# Improving the performance of Bandwidth Efficient Acknowledgement based Multicast (BEAM) protocol in VANET for Urban environment

Alehegn Minale [1,2*], Dawit Keflie [2,3†] and Getamesay Haile[1,2†]

[1*] Information Technology, Jimma University, Jimma, Oromia, Ethiopia.

[2] Computer Science, Chicago, USA.

[3] Information Technology, Griffith University, Queensland, Austria.

*Corresponding author(s). E-mail(s): alixminale@gmail.com;

Contributing authors: kifledawit@gmail.com; getamesay0923@gmail.com;

†These authors contributed equally to this work.

## Abstract

*Vehicular Ad-hoc Network (VANET) is part of Mobile Ad-hoc Network (MANET) that facilitates communication between group of vehicles to provide safety of vehicles and road, traffic updates, entertainment, data sharing and also to increase the comfort to both drivers and passengers. Due to the high mobility and dynamism, routing the messages/disseminate the data to their final destination in VANETs is a challenging task. There are number of protocols that disseminate data to appropriate destination in VANET but in order to preserve limited bandwidth and to disseminate maximum data over networks during emergency situation makes it more challenging. These protocols are relegated as broadcasting, multicasting and geocasting based protocols. Multicasting based protocols are found to be best for conserving the bandwidth. The one and best desired protocol named as Bandwidth Efficient Acknowledgment Based Multicasting Protocol (BEAM) exists that improve the performance of VANETS by repressing the number of in-network message transactions and thereby efficiently using the bandwidth during an emergency situation. But this protocol may leads to multicar chain collision as there is no V2V communication. So in order to overcome the problem of BEAM protocol we proposed algorithm by considering the concept of clustering. Clustering is a way of grouping of vehicles based upon some predefined metrics such as density, velocity, and geographical locations of the vehicles. Clustering in vehicular ad hoc networks (VANET) is one of the control means for dynamic topology. We established stable clustering based on speed and direction of vehicles to provide stability, which in turn enhance cluster life time and minimize communication barriers and installation cost of RSU. Proposed approach make use of multi-agent interaction method consists of static and mobile agents to establish the communication between vehicles and RSU. RSU for multicast groups whereas CH(in MG) for non-multicast group agents are responsible to decide the cluster size and select suitable primary and secondary cluster head based on vehicle speed and connectivity with neighboring nodes. The second cluster head used as backup when the first cluster head is failed and act as main cluster head to prevent re-clustering in high mobility and dynamism of vehicles over the network. The cluster size is vary depending upon vehicle speed, which in turn increases the cluster life time and reduces the routing overhead.Simulation results show the improved performance of new proposed algorithm in terms of Throughput, PDR and end to end delay when compared with BEAM protocol.*

*Keywords: Vehicular Adhoc network, MANET, BEAM, CH, RSU, MG, Emergency situation, V2V*

# 1. Introduction

Vehicular Ad Hoc Networks (VANETs) are a specialized form of Mobile Ad Hoc Networks (MANETs), where mobile nodes are vehicles. They facilitate wireless communication between vehicles and roadside equipment [1], [2], [3], [5]. VANETs are distributed, self-organized, and potentially highly mobile networks of vehicles interacting via wireless media [4]. In these networks, the high mobility of vehicles results in each vehicle acting as both a router and a host, leading to rapid changes in network topology [6], [7].

The topology in VANETs frequently changes, and vehicles are not always connected to the network, resulting in intermittent disconnections. There are four vital characteristics that differentiate MANETs from VANETs:

1. High mobility of nodes, leading to rapid topology changes.
2. Unstable wireless channel quality influenced by factors like roadside infrastructure and vehicle velocity.
3. Utilization of GPS and digital maps for precise positioning and clock synchronization.
4. Predictable vehicle movement due to static road structures [8].

These factors suggest that traditional routing protocols used in MANETs may not be suitable for VANETs, as they often assume continuous network connectivity [6], [7]. Moreover, the intermittent connectivity, frequent changes in network topology, and low reception rates are distinct properties that set VANETs apart from other types of ad hoc networks [6]. While VANETs share similarities with ad hoc networks, they also possess unique characteristics that require tailored solutions [8].

Routing in VANETs involves searching for an optimal path between source and destination nodes to ensure messages reach their destinations efficiently and timely. A significant challenge in VANETs is how to exchange information in a scalable manner [9]. This is where Data Dissemination Protocols come into play. These protocols vary based on their applicability to different scenarios, such as highways versus urban areas.

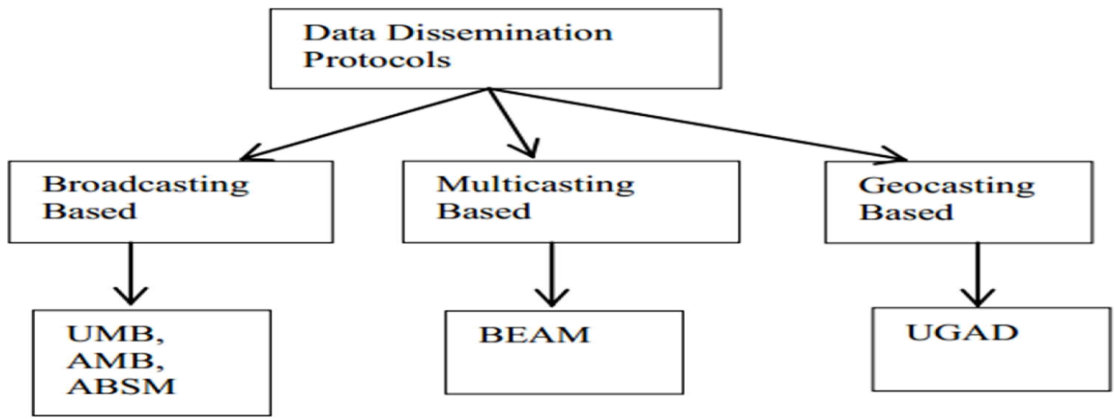

Figure 1 Data dissemination protocol [8]

This study specifically focuses on multicasting-based protocols. According to [11], multicasting is advantageous for conserving bandwidth. The Bandwidth Efficient Acknowledgment based Multicasting protocol (BEAM) improves VANET performance by reducing in-network message transactions, thereby efficiently utilizing bandwidth during emergency situations. BEAM employs multicasting, transmitting messages only to vehicles within its transmission range rather than broadcasting to all [8]. In this protocol, emergency situations are predicted based on status reports sent by vehicles to nearby Road Side Units (RSUs), which create and send emergency messages to all group members, who then reply with acknowledgments.

However, this approach may lead to multicar chain collisions since it does not facilitate vehicle-to-vehicle (V2V) communication, leaving ordinary vehicles uninformed about emergencies. To address this issue, researchers proposed the Enhanced Bandwidth Efficient Cluster based Multicasting protocol (EBECM), which utilizes clustering to facilitate V2V communication and inform non-multicast group members (ordinary vehicles) about emergencies within the communication range. Vehicles can decide whether to accept messages, with interested vehicles replying to join control packets sent to the RSU. Yet, this protocol is limited to vehicles within the RSU's communication range, necessitating multiple RSUs for full coverage, which poses significant cost and logistical challenges [30], [31].

The motivation behind this study is the alarming number of fatalities caused by communication barriers between vehicles, emphasizing the need for improved bandwidth utilization during emergencies while minimizing RSU installation costs. To address the limitations of the BEAM protocol, this study proposes a clustering algorithm that enhances the performance of the Bandwidth Efficient Acknowledgment based Multicasting protocol by forming clusters for both multicast and non-multicast vehicles.

VANETs have attracted significant interest from researchers in academia and industry due to their potential applications in various Intelligent Transportation Systems (ITS) that enhance safety, entertainment, emergency response, and content sharing [10]. Despite prior studies on optimizing bandwidth utilization during data dissemination, existing protocols often neglect ordinary vehicles outside multicast groups [11]. This oversight can lead to multicar chain collisions, as there is no effective V2V communication to inform non-multicast members of emergencies.

While EBECM attempts to resolve these issues by fostering V2V communication through clustering, it still confines communication to vehicles within the RSU's range. The high cost and limited deployment of RSUs necessitate innovative solutions that extend communication capabilities to all vehicles in the network, including those outside the RSU's transmission range.

This paper aims to explore improved cluster formation within the BEAM framework to reduce packet drop rates, minimize end-to-end delays, and maximize throughput compared to existing protocols.

The study of Vehicular Ad Hoc Networks (VANETs) is significant as it provides a framework for intelligent transportation systems, where vehicle nodes function both as routers and hosts to propagate critical information among nearby vehicles and roadside equipment, thereby enhancing safety, travel time, and speed [1]. Efficient data dissemination is crucial for preventing collisions, accidents, and traffic jams, enabling various applications such as traffic reporting, emergency warnings, and road condition monitoring. This research proposes an algorithm that improves the performance of the Bandwidth Efficient Acknowledgment based Multicast protocol by ensuring efficient bandwidth utilization during emergencies, thereby saving lives, minimizing collisions, and reducing the installation costs of roadside units. The methodology involves experimental research with simulations conducted using various tools, including Ns2 and SUMO, to evaluate clustering mechanisms based on node speed, position, and communication range. The ultimate goal is to optimize communication efficiency and support vehicle-to-vehicle interactions both within and beyond the communication range [2], [3]. This paper comprises introduces the study's highlights, including the problem statement, objectives, scope, significance, and methodology and literature Review provides a detailed description of data dissemination routing protocols, focusing on clustering formation algorithms and the BEAM protocol's characteristics and advantages. It also discusses the implemented techniques and the new algorithm for cluster head selection and maintenance, and the outlines the proposed solution using existing techniques and protocols. The next part the paper presents the implementation of the modified protocol along with simulation results, analysis, and interpretation. Finally, concludes the study and suggests directions for future work.

**Literature Review**

Ad hoc networks are decentralized communication systems comprised of numerous devices that operate without a fixed infrastructure, relying on nodes to dynamically discover neighbors for communication. These networks can be categorized into Wireless Mesh Networks (WMN), Mobile Ad hoc Networks (MANET), and Wireless Sensor Networks (WSN) [17]. MANETs, which include Vehicular Ad-hoc Networks (VANET), Intelligent Vehicular Ad-hoc Networks (InVANETs), and Internet-Based Mobile Ad-hoc Networks (iMANET), play a crucial role in modern transportation systems as illustrated in the figure 2 below.

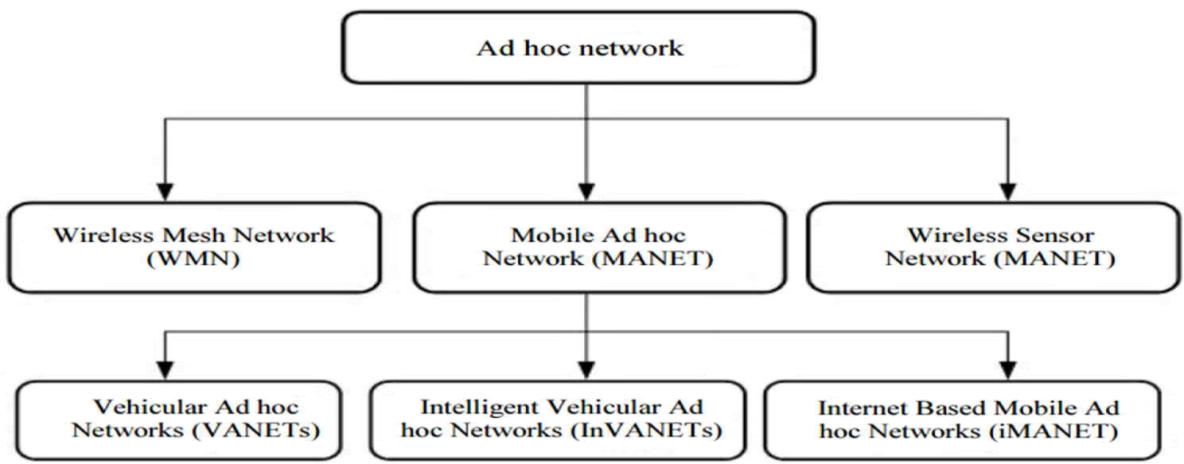

Figure 2  Ad hoc networks categories [17].

VANET utilizes vehicles as nodes to form a mobile network, enabling direct communication between vehicles (Vehicle-to-Vehicle, V2V) and between vehicles and roadside infrastructure (Vehicle-to-Infrastructure, V2I) [17]. The architecture supports various applications and operates using Dedicated Short Range Communication (DSRC) technology within a designated radio spectrum [23].

Key characteristics of VANET include high mobility of nodes, dynamic network topology, and frequent information exchange [17]. Applications of VANET are categorized into safety-oriented (e.g., real-time traffic monitoring), commercial (e.g., internet access), and convenience-oriented (e.g., traffic management) [18]. These features and applications underscore the potential of VANET in enhancing road safety and traffic efficiency.

Data Dissemination in VANET

Data dissemination refers to the process of spreading information across distributed wireless networks, aiming for optimal use of network resources to meet users' data needs [8]. In VANETs, several types of data dissemination methods are employed, including Vehicle-to-Infrastructure (V2I), Vehicle-to-Vehicle (V2V), Opportunistic Dissemination, Peer-to-Peer (P2P) Dissemination, and Cluster-Based Dissemination.

| *Dissemination Type* | *Approach* | *Pros* | *Cons* |
|---|---|---|---|
| **V2I (Vehicle to Infrastructure)** | Push-based | Suitable for popular data | Unsuitable for non-popular data |
| | Pull-based | Suitable for non-popular, user-specified data | Cross traffic incurs heavy interference and collisions |
| **V2V (Vehicle to Vehicle)** | Flooding | Quick data distribution and reliability | Not suitable for dense networks |
| | Relaying | Effective in dense and congested networks | Difficulty in selecting the next hop and ensuring reliability |
| **Opportunistic** | Store and forward | Dynamically built routes | Data-centric architecture may lack efficiency |
| **P2P (Peer-to-Peer)** | Store and forward on demand | Effective in delay-tolerant applications | Messages may not propagate throughout the network |
| **Cluster-Based Dissemination** | Clusters are generated | High delivery ratio with low delay | Limited broadcasting capability for all nodes |

Table 1: Comparison of Data Dissemination Types) [8]

**Overview of Data Dissemination Protocols**

Various protocols exist in VANET for data dissemination, including broadcasting, geo-casting, and multicasting, each with its advantages and disadvantages as illustrated in Figure 1 above.

**Broadcast-Based Protocol in VANET:-**Broadcast routing is commonly used in VANET for sharing traffic, weather updates, and emergency information among vehicles. This method allows messages to reach vehicles beyond the transmission range through multi-hop connections [22]. However, broadcasting with acknowledgments can increase in-network traffic, leading to inefficient bandwidth usage [11].

**Geo-Casting Based Protocol in VANET**: - Geo-casting involves delivering messages to nodes within a specified geographical area. This approach enables new services such as locating nearby friends, geographic advertising, and issuing warnings for accidents or wrong-way drivers [21].

**Multicast-Based Protocol in VANET: -** Multicast protocols enable messages to be sent from a single source to multiple destinations within a communication range [19]. Vehicles receive messages based on their interests, reducing unnecessary traffic and optimizing bandwidth usage. For example, the Bandwidth Efficient Acknowledgment Based Multicasting (BEAM) Protocol minimizes in-network message transmission during emergencies [11]. The BEAM protocol typically employs Roadside Units (RSUs) to facilitate multicast group establishment. The RSU broadcasts join control packets, and vehicles entering its range can opt into receiving emergency messages by replying with control packets. BEAM includes modules for dynamic group establishment, emergency situation prediction, member verification, and acknowledgment for emergency messages. However, it does not support vehicle-to-vehicle communication, which may lead to chain collisions among multiple vehicles [11].

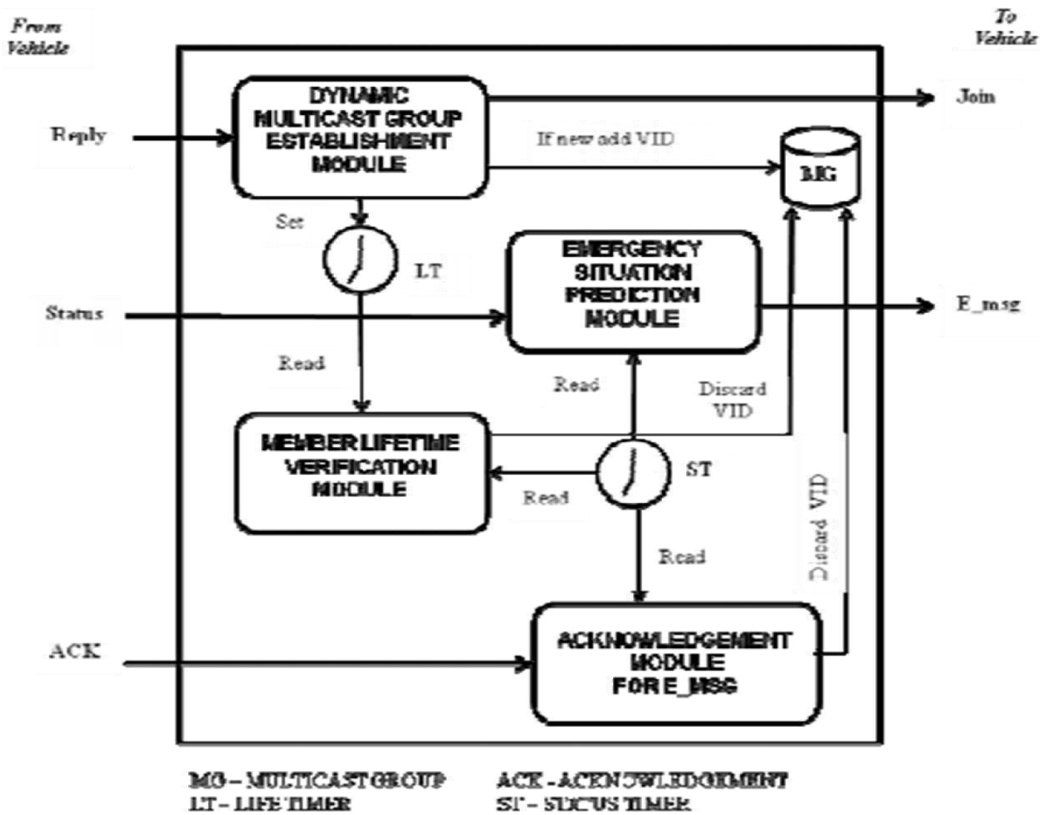

Figure 3: General Architecture of BEAM Protocols) [11]

**Wireless Access in Vehicular Ad-hoc Networks (VANETs)**

The development of products within vehicular networks is significantly streamlined by established standards, which facilitate product development, reduce costs, and enable user comparisons among competing offerings. Interconnectivity and interoperability are predominantly achieved through these standards, necessitating the verification of new products to ensure the swift integration of emerging technologies. Various standards govern wireless access in vehicular environments, notably the Dedicated Short Range Communication (DSRC) and the IEEE 1609 standards for Wireless Access in Vehicular Environments (WAVE) [24]. DSRC is tailored for short to medium-range communications, supporting Vehicle-to-Vehicle (V2V), Vehicle-to-Roadside (V2R), and hybrid communications, facilitating applications such as safety messaging and toll collection with minimal latency and high reliability [24]. Furthermore, WAVE, which utilizes existing IEEE 802.11a compliant devices, addresses the challenges presented by dynamic vehicular scenarios, providing a framework for Road Side Units (RSUs) and Onboard Units (OBUs) to interact effectively [34]. The deployment of RSUs in urban settings is crucial for enhancing message dissemination reliability; however, the associated costs and optimal placement strategies are critical considerations [35]. Moreover, clustering within VANETs enhances routing efficiency, particularly in the context of high mobility and dynamic network conditions, by organizing vehicles into sub-networks based on metrics such as density and velocity, thereby improving overall network stability [16, 35].

**Related Works: -** This study is supported by research in VANETs aimed at improving data transmission accuracy and driver safety through various dissemination protocols.

The **Urban Multi-hop Broadcast (UMB)** protocol tackles broadcast storms and hidden nodes in urban areas through directional broadcasting but incurs costs for repeaters [12].

The **Ad-hoc Multi-hop Broadcast (AMB)** protocol extends UMB without repeaters, allowing vehicles to choose the closest vehicle to intersections for packet forwarding, though it may waste time in the selection process [13].

**Acknowledgment-Based Broadcast from Static to Highly Mobile Protocol (ABSM)** allows vehicles to decide whether to forward packets based on local information, reducing redundancy but potentially causing high overhead during simultaneous broadcasts [14]

The **Urban Geo-cast based on Adaptive Delay (UGAD)** protocol focuses on delivering messages to visible vehicles at intersections, improving delivery ratios but lacking delivery confirmation [15].

**Bandwidth Efficient Acknowledgment Based Multicasting Protocol (BEAM)** uses RSUs to efficiently transmit messages only to interested vehicles but risks multicar collisions due to lack of V2V communication [11].

Lastly, the **Enhanced Bandwidth Efficient Cluster Based Multicasting Protocol (EBECM)** promotes V2V communication within clusters but does not address vehicles outside RSU range, leading to higher costs and reduced effectiveness in preventing collisions [8].

**Clustering Algorithm in VANET**

Clustering in vehicular ad hoc networks (VANETs) serves as a critical control mechanism to stabilize the network topology, thereby alleviating the load on central control units and reducing data redundancy [38]. Many of the clustering algorithms used in VANETs are derived from mobile ad hoc networks (MANETs). However, these algorithms often overlook the unique mobility characteristics of VANET nodes, which are characterized by high speeds and unpredictable movement patterns. This paper does not introduce new clustering algorithms but leverages existing ones to achieve efficient and stable clustering tailored to VANET dynamics.

**Lowest ID (LID)** is a straightforward algorithm that selects the node with the lowest ID as the Cluster Head (CH), thereby facilitating identifier-based clustering [26]. This approach helps maintain a hierarchy within clusters but may struggle with nodes that frequently change roles due to mobility.

The **Highest Degree (HD)** algorithm, on the other hand, focuses on connectivity, selecting the node with the most connections as the CH. This enhances the cluster's ability to relay information effectively [27]. Nodes within transmission range of this CH receive their IDs and rely on the CH for communication, which optimizes information exchange but may lead to rapid changes in leadership as nodes move.

**Low Energy Adaptive Clustering Hierarchy (LEACH)** promotes energy efficiency by selecting CHs through a randomized process during each round of communication. This self-organized and self-adaptive protocol consists of a setup stage and a steady state, allowing for energy conservation among nodes [25]. The alternating CH roles in LEACH help distribute energy usage, prolonging the network's overall lifetime.

The **Direction-Based Clustering Algorithm (DCA)** uses directional metrics to classify neighbors, identifying close neighbors and further enhancing cluster stability by electing a CH based on proximity [25]. This method considers the relative motion of vehicles, ensuring that CHs remain relevant to the cluster's dynamics.

The **Cuckoo Search (CS)** optimization algorithm draws inspiration from the reproductive behavior of cuckoos to select a super cluster head (SCH). This approach optimizes network performance by identifying the best candidate among multiple CHs, focusing on metrics like network lifetime and packet delivery ratio [25]. The SCH plays a crucial role in maintaining efficient communication among clusters.

**Mobility-Aware Clustering** focuses on enhancing stability by considering node mobility, neighbor count, and cluster lifetime. This algorithm operates in two phases: setup and maintenance, assigning a primary cluster head (PCH) and a secondary cluster head (SCH). If the PCH moves out of range, the SCH takes over, ensuring continuity in leadership [28].

The **Traffic Flow-Based Algorithm** establishes stable clusters by selecting CHs based on traffic flow metrics. Vehicles calculate a cluster head leave (CHL) considering network connectivity, average distance, and average velocity, focusing on lanes with significant traffic [28]. This approach improves cluster stability and prolongs its lifetime by ensuring that the CH is optimally positioned.

**Combined-Metrics-Based Clustering** algorithm employs a force-directed mechanism, where nodes exert forces on each other based on relative velocity and distance. This technique fosters cluster formation by encouraging nodes moving in the same direction to group together, while those moving oppositely are less likely to form clusters [28]. The algorithm calculates total forces to determine the likelihood of a node becoming a cluster head, promoting both stability and efficiency in dynamic environments.

**Cluster Formation**

According to IEEE 802.11p Wireless Access in Vehicular Environments (WAVE), two device types are defined: Road Side Units (RSUs) for fixed communication and Onboard Units (OBUs) for vehicles. OBUs act as static agents, storing their location and information about neighboring vehicles, including vehicle ID, speed, and direction [24]. This paper used SUMO tools to analyze real traffic mobility data.

In this paper we have considered new algorithm to make clustering, for multicast groups, the RSU determines cluster size and selects primary and secondary cluster heads based on vehicle speed and connectivity. In non-multicast groups, the cluster head from the multicast group selects two cluster heads and the cluster size. Varying the cluster size based on vehicle speed enhances cluster lifetime and reduces routing overhead [24].

**Mathematical Modeling**

Let C1, C2, C3… Cn represent vehicles moving in a single-lane system with variable speeds. The RSU and Cluster Head (CH) for non-multicast groups collect information about vehicle ID, speed, neighbor lists, and positions. To determine whether a vehicle is within the communication range, the distance to the RSU is calculated. This study considers two fixed infrastructures (RSUs) equipped with GPS [24].

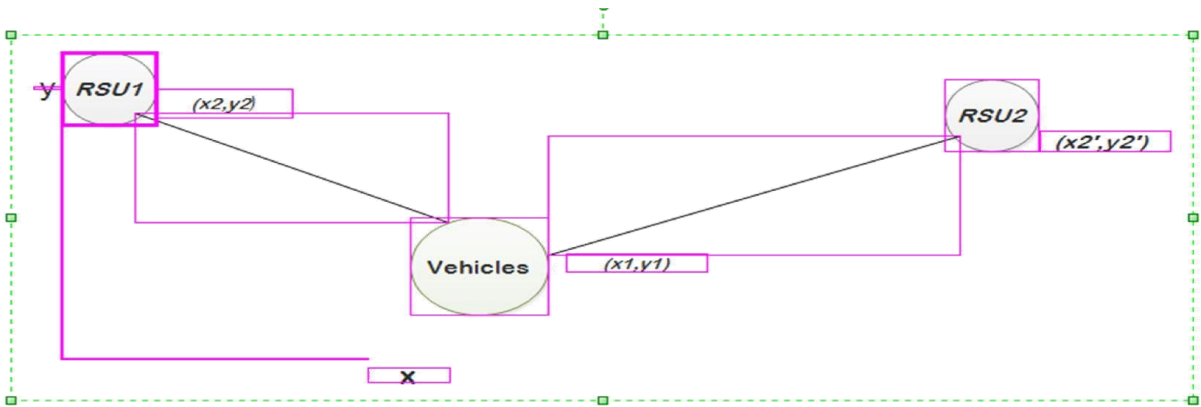

Figure 4. Distance calculation in between two Roadside units and vehicles [24]

$$Distance = \sqrt{(x2' - x2)*(x2' - x2) + (y2' - y2)(y2' - y2)} \quad \text{............(1)}$$

The RSUs, fixed and equipped with GPS, can utilize arbitrary coordinates. In this paper, RSU1 is located at (200, 200), and RSU2 is deployed at (1200, 200), maintaining a deployment distance of approximately 1000 meters [35]. The communication range for each vehicle and RSU is around 300 meters.

Now let's calculate the distance of the vehicles relative to RSU1 and RSU2

$$D1 = \sqrt{(x2 - x1)*(x2 - x1) + (y2 - y1)(y2 - y1)} \quad \text{............(2)}$$

$$D2 = \sqrt{(x2' - x1)*(x2' - x1) + (y2' - y1)*(y2' - y1))} \quad \text{............(3)}$$

Then to calculate the average speed first we should calculate speed of each vehicle relative to a certain time interval.

$$\text{Speed} = \frac{\text{Distance}}{\text{Time}} \quad \text{............(4)}$$

Initially, vehicle V is located at coordinates (x1, y1) at time t1. After a period until time t2, the vehicle moves to a new position at (x2, y2), the minimum time interval for position change in NS2 is 30 seconds. Therefore, the speed of each vehicle can be calculated based on the change in position between these two points over the time interval from t1 to t2.

- t1:- Initial position time (constant).

- t2:- Time when the vehicle reaches the new position.

$$\Delta t = t2 - t2 \quad ................................................................................(5)$$

Then to calculate the speed of the vehicle lets us consider equation 2 and equation 5

$$speed = \sqrt{(x2-x1)*(x2-x1)+(y2-y1)(y2-y1)}/\Delta t \quad ........................ (6)$$

Then after calculating the speed of each vehicles, computes average speed of vehicles is given in equation (7).

$$\text{Savg} = \sum_{i=1}^{n} \frac{si}{\rho} \quad ............................................................... (7)$$

Where Savg= average speed, $S_i = s_1, s_2, s_3, S_4 \ldots S_n$, $\rho$= Vehicle density (number of vehicles)

Cluster size is directly related to vehicle density and average speed. The radius of a cluster varies dynamically based on the vehicles' average mobility and density [29]. Additionally, as noted in reference [32], the relative direction of two vehicles can be determined by the angle θ between their velocity vectors. Let's consider the positions of the two vehicles v1 and v2 at time t are (x1, y1), (x2, y2), and at time t + Δt (where Δt is a short time) are (x1^, y1^), (x2^, y2^), respectively, as shown in Figure 5 below.

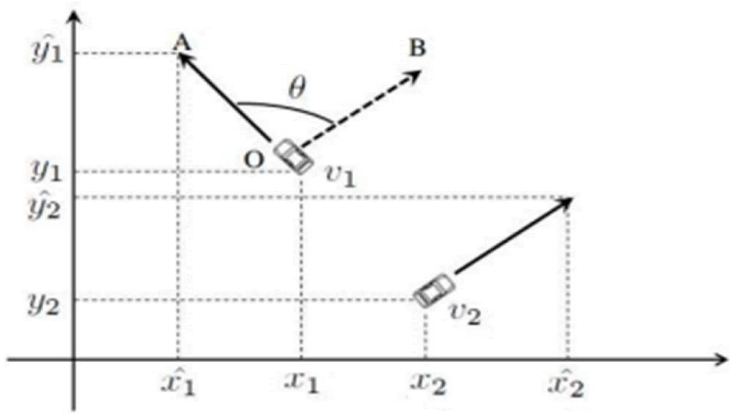

Figure 5 Moving direction angle calculation [32]

The angle θ between two given velocity vectors is given by the following expression [33]

$$\overrightarrow{OA}.\overrightarrow{OB} = \overrightarrow{||OA||}\;\overrightarrow{||OB||} * cos\theta \quad ............................................................... (8)$$

$$\theta = arcos(\frac{\Delta x1*\Delta x2+ \Delta y1*\Delta y2}{\sqrt{(\Delta x1^2+\Delta x2^2)}\,(\sqrt{\Delta y1^2+ \Delta y2^2})}) ............................................... (9)$$

After receiving a HELLO message from each of its one-hop neighbors, vehicle (i) only considers those neighbors whose angular directions fall within to $\theta_i \pm \delta$, where $\theta_i$ is the direction of vehicle (i) and $\delta$ represents the acceptable range of angles for vehicles moving in the same direction. According to the authors in reference [34], two velocity vectors are considered non-parallel if the smallest angle between them exceeds 18 degrees. Consequently, vehicle (i) disregards HELLO messages from neighbors with non-parallel velocity vectors. This directional analysis aids in forming a stable cluster structure by grouping vehicles with parallel velocity vectors together. Cluster head selection has been considered with a cluster of vehicles, one vehicle is chosen as the primary cluster head, while another is selected as the secondary cluster head. This selection is based on factors such as vehicle speed, direction, and neighbor connectivity. To determine the most suitable primary and secondary cluster heads, a weight factor (WT) is calculated [29].

$$WT \propto \frac{NC}{Snavg} \quad \ldots\ldots\ldots\ldots\ldots\ldots\ldots\ldots\ldots\ldots\ldots\ldots\ldots\ldots\ldots\ldots (10)$$

Where $N_c$ = Neighbors Count, $S_{navg}$ = near to above average speed

After cluster formation, each vehicle transmits its weight factor (WT) to Roadside Units (RSUs) and Cluster Heads (CHs) to facilitate Vehicle-to-Vehicle communication. RSUs and CHs select the most appropriate primary and secondary cluster heads based on three criteria: near to above average speed ($S_{navg}$), direction, and neighbor count ($N_C$). Vehicles with above-average speed and a higher number of neighboring nodes are chosen as the primary CH. To enhance cluster stability and minimize re-clustering, the vehicle with the second highest weight factor is designated as the secondary cluster head. In the event of a primary CH failure, control shifts to the secondary CH, allowing it to take over as the primary head and reducing the need for re-clustering in the dynamic context of VANET networks. The high dynamic nature of VANETs leads to frequent joining and leaving of vehicles within clusters, resulting in additional maintenance overhead. The events triggering maintenance can be summarized as follows:

1. If a new node enters to the communication range of an existing cluster, the Cluster Head (CH) adds it to the cluster member (CM) list.
2. If an existing CM exits the communication range of the CH, the CH removes it from the CM list.
3. If the current CH fails, the secondary cluster head is promoted to CH.

**Proposed Solution**

The goal of this work is to enhance the Bandwidth Efficient Acknowledgment Based Multicast protocol by clustering ordinary vehicles (those outside RSU communication range) and multicast group vehicles to support Vehicle-to-Vehicle (V2V) communication and reduce RSU deployment costs. The proposed architecture aims to create separate multicast and non-multicast groups to optimize limited bandwidth in VANETs during emergencies, thereby minimizing packet drop rates and delays while increasing throughput. The source node broadcasts a join control packet to nearby vehicles to form the multicast group, while the non-multicast group is identified by a cluster head selected from the multicast group.

Current protocols lack V2V communication capabilities for ordinary vehicles during emergencies, and leaving them unable to notify RSUs or multicast groups. To address this, clustering concepts are employed to bridge communication gaps. the distance between vehicles and RSUs is assessed above on this paper, classifying vehicles within RSU range and responding to the join control packet as the multicast group, while those outside this range are designated as the non-multicast group. The proposed architectural procedure is illustrated in Figure 6 below.

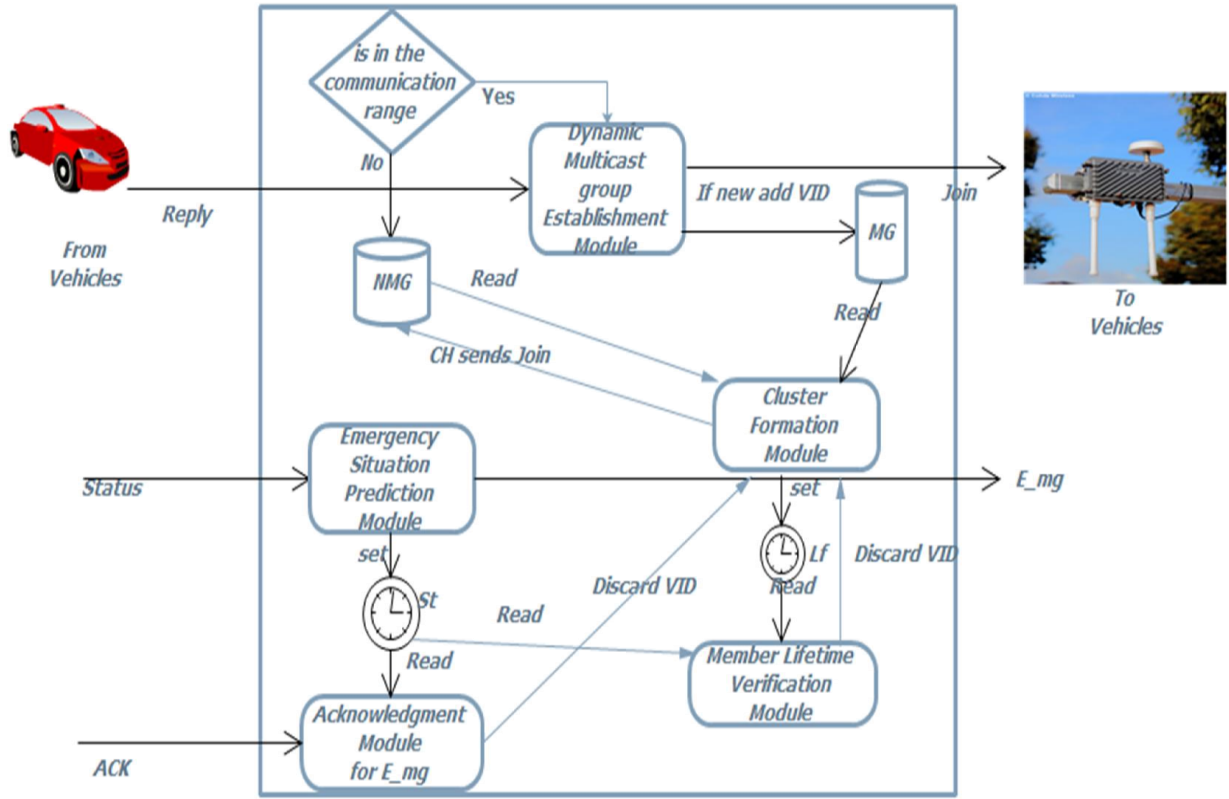

Figure 6  Modified Architecture of BEAM

The proposed solution aimed to enhance the Bandwidth Efficient Acknowledgment Based Multicast protocol by clustering ordinary vehicles (those outside RSU communication range) and multicast group vehicles to improve Vehicle-to-Vehicle (V2V) communication and reduce RSU deployment costs. Key Components

1. **Emergency Situation Detection**: Vehicles exhibiting abnormal parameters, such as a sudden increase in speed or yaw rate, initiate an abnormal status report to RSUs and Cluster Heads (CHs). If significant changes are detected, an emergency message (E_mg) is generated and disseminated to all relevant parties.
2. **Communication Architecture**: The protocol is divided into two main components: one for RSUs and another for vehicles. The RSU protocol comprises five modules:
    - **Dynamic Multicast Group Establishment**: Vehicles within RSU communication range can join a multicast group through periodic broadcasts.
    - **Cluster Formation**: This module forms clusters based on vehicle speed, direction, and neighbor connectivity, setting a lifetime for each member.
    - **Emergency Situation Prediction**: Monitors vehicle status and triggers emergency messaging if abnormal conditions are detected.
    - **Member Lifetime Verification**: Ensures vehicles remain active in the cluster; inactive vehicles are removed after timeouts.
    - **Acknowledgment for Emergency Messages**: Implements an acknowledgment mechanism to confirm receipt of emergency messages.
3. **Vehicle Protocol**: This consists of three modules:
    - **Cluster Formation**: Similar to the RSU protocol, it organizes vehicles into clusters based on direction and speed.
    - **Response Controller**: Facilitates communication between vehicles and RSUs/CHs, ensuring timely acknowledgment of emergency messages.
    - **Sense and Report**: Vehicles report their current status, including speed and yaw rate, to RSUs and CHs at scheduled intervals.

The proposed architecture aims to minimize packet drop rates and delays, increasing throughput and enhancing the reliability of emergency communications in dynamic VANET environments. In vehicular Ad-hoc Networks (VANETs), efficient routing of information is crucial to prevent collisions, accidents, and traffic jams. Current data dissemination protocols, particularly the BEAM protocol, focus on multicast communication through Roadside Units (RSUs) to manage bandwidth during emergencies. However, BEAM does not facilitate communication from ordinary vehicles outside the multicast group, which poses a risk during emergencies where information needs to be shared quickly to prevent accidents.

To address this gap, the proposed Enhanced Bandwidth Efficient Cluster-based Multicast Protocol (EBECM) allows for vehicle-to-vehicle communication, incorporating ordinary vehicles into the network. The RSU sends join control packets to all nearby vehicles to form a multicast group based on their response. Vehicles that do not respond are classified as non-multicast (ordinary vehicles). The algorithm operates as follows:

1. **Join Control Packet:** RSUs send a packet to vehicles within their communication range.
2. **Multicast Group Formation:** Vehicles that respond are added to the multicast group, allowing RSU communication only among these vehicles.
3. **Cluster Formation:** The RSU evaluates vehicle speed and direction to establish clusters within the multicast group, selecting two cluster heads based on a weighting factor to maintain stability.
4. **Communication with Non-Multicast Vehicles:** Selected cluster heads broadcast join control packets to nearby ordinary vehicles, facilitating communication beyond the RSU's range.
5. **Emergency Reporting:** In emergencies, ordinary vehicles report to their nearest cluster head, which then relays information through the RSU to ensure rapid dissemination to all cluster members.

This algorithm reduces the need for numerous RSUs by strategically installing them in high-traffic areas, spaced 1000 meters apart. It enhances communication during emergencies, allowing ordinary vehicles to participate in critical information exchange, thus addressing potential multi-car collisions effectively.

**Algorithm 1: Pseudo Code of proposed algorithm:**

| | |
|---|---|
| Initialization | *S = Source (i.e. RSU)*, **MG** = *Set of multicast group members.* <br> **NMG**=*Set of Non-multicast group member , C= Communication range* <br> **Savg**=*Average speed of vehicle ,***Sth**= *Threshold speed of vehicle* <br> **CH**=*Cluster Head , ***SH**= *Second cluster head* <br> **CM**=*Cluster Member ***WT**=*Weight Factor ***Ov**=*Other Vehicle* <br> **N**= *number of Vehicle ***cspeed**= *current Speed ***rspeed** =*Reported Speed* <br> **cyaw_rate** = *Current yaw_rate ## current position of Vehicle,* **ryaw_rate** =*Reported yaw_rate,***d1** = *distance one,* **d2** = *distance two* |
| Definition | *VID – Vehicle ID.* <br> *E_msg – Emergency Message.* <br> *Periodic_timer – Set as 1 second.* ### **Time used to send Join control packet** <br> *Status_timer – Set as 1 second.* #### **Time used to predicate emergency** <br> *Life_timer – Set as 30 seconds.* ### **Life time of Vehicle in the cluster** <br> *Ack_timer – Acknowledgement timer (Set as 1 second).* |
| 1. *Sleep until periodic_ timer mature*<br>*S ← broadcast control join packet*<br>2. *If ((((d1< C) && (d2> C)) || ((d1> C) && (d2< C))) &&(Vehicles reply control join packet in the communication range))*<br>*Then S Extract VID and check the VID in MG*<br>*If (Found)*<br>*Discard VID Else*<br><br>*Add the VID to MG End if* | 3. *S get basic information includes VID, position (xi, yi), directions, speed, negibors_list, density, distance ##to make Cluster*<br>*If (θ < = 18$_0$)*<br>*For ← i =1 to Val (N) each vehicle forwards speed If S$_{avg}$ > S$_{th}$ then*<br>*S forms large cluster members then Set Life_timer Else*<br>*S forms small cluster members then set Life_timer End if*<br>*For ← i=1 to Val (N) ### Cluster head Selection If WT is highest*<br>*Select vehicle as main CH Else if WT is second highest value*<br>*Select vehicle as SCH Else*<br>*Select Vehicle as CM End if*<br>*If Ov enters into C of S ### Cluster maintenance S adds Ov to CM list ##*<br>*Else if CM exits from C || Life_timer expire S deletes CM*<br>*Else if CH fails to work ## out of C |Life_timer expire SCH will be CH*<br>*End if End if* |
| 4. *If (cspeed> (rspeed + (1/3) * rspeed)) || (cyaw_rate> (ryaw_rate+ 30 degrees))*<br>*S ← broadcast E_msg to CH, CH ← CM&&set Ack_timer*<br>*If (Vehicle received E_msg)*<br>*Send ACK ### to the corresponding source*<br>*Else check the VID in cluster member If (found) Increment attempt value by 1 then If (Attempt greater than maximum)*<br>*Discard the VID from the member list Else*<br>*Resend E_msg to CH, CH ← CM End if*<br>*End if End if* | 5. *Else ((d1> C) && (d2>C)) && (VID not in MG) ## i.e. node consider as NMG*<br>*CHs (MG) ← broadcast control join packet If (Vehicles reply join control packet) then*<br>*CHs (MG) used as S (RSU) for NMG and goto Step 3. Else exit*<br>*End if End if* |

Table 2: Pseudo Code of proposed algorithm

In the pseudo code, the source node (S) initiates communication by sending a join control packet to establish a Multicast Group (MG). Cluster Heads (CHs) within this group then relay join control packets to vehicles outside the source's communication range, facilitating broader information distribution. The roles are clearly defined: S serves the Multicast Group, while CHs manage communication with the Non-Multicast Group, ensuring stability in the dynamic VANET environment. For emergency situations, the RSU sends warning messages to the Multicast Group, while the CHs handle status reports from vehicles to disseminate alerts to ordinary vehicles. This structure maximizes information sharing across all vehicles on the road.

**Implementation and evaluation**

The proposed algorithm using NS-2 and SUMO-0.26.0 tools, focusing on minimizing packet drop rates, reducing end-to-end delays, and enhancing throughput. The paper assumes that vehicles operate at varying speeds influenced by road conditions and traffic, with an average threshold speed derived from typical expressway limits in India. Real-time traffic scenarios were modeled using data from OpenStreetMap, specifically at Kathipara Junction in Chennai, due to the lack of suitable urban expressways in Ethiopia. Traffic mobility data generated in SUMO was exported to NS-2 for performance analysis, validating the emergency prediction scheme and the efficacy of the cluster-based data dissemination protocol in realistic traffic conditions.

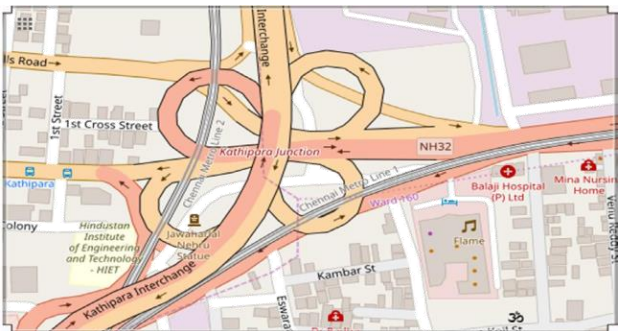 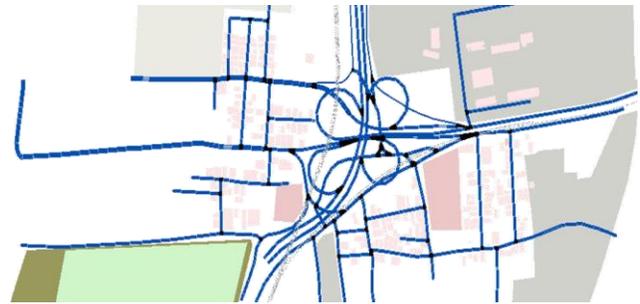

Figure 7 Map of Tested Area    figure 8: Network topology by using SUMO

| Parameters | Values |
| --- | --- |
| Channel | Wireless |
| Propagation Model | Two Ray Ground |
| MAC layer | 802.11p and IEEE1609.4 |
| Physical layer | 802.11p |
| Number of vehicle | 25 |
| Number of RSU | 2 |
| Simulation time | 500sec |
| Transmission range | 300m |
| Simulation area | 2850.46 x 2000.04 |
| Speed of Vehicles | 80-120km/h |
| Threshold speed of vehicles | 100km/h |

**Table 3 Simulation Parameters**

**Performance Evaluation Metrics and Parameters**

To assess our proposed algorithm and to compare it with the existing BEAM data dissemination protocol, we have utilized various performance metrics. These metrics included throughput, end-to-end delay, and packet delivery ratio.

**Throughput**: It is defined as number of packets that are passing through the channel in a particular unit of time. [20]

$$Throughput(kbps) = \frac{(Total\ Recieved\ packets\ Size)*(8)}{StopTime - StartTime} \quad \dots\dots\dots\dots\dots\dots\dots\dots\dots\dots\dots\dots(11)$$

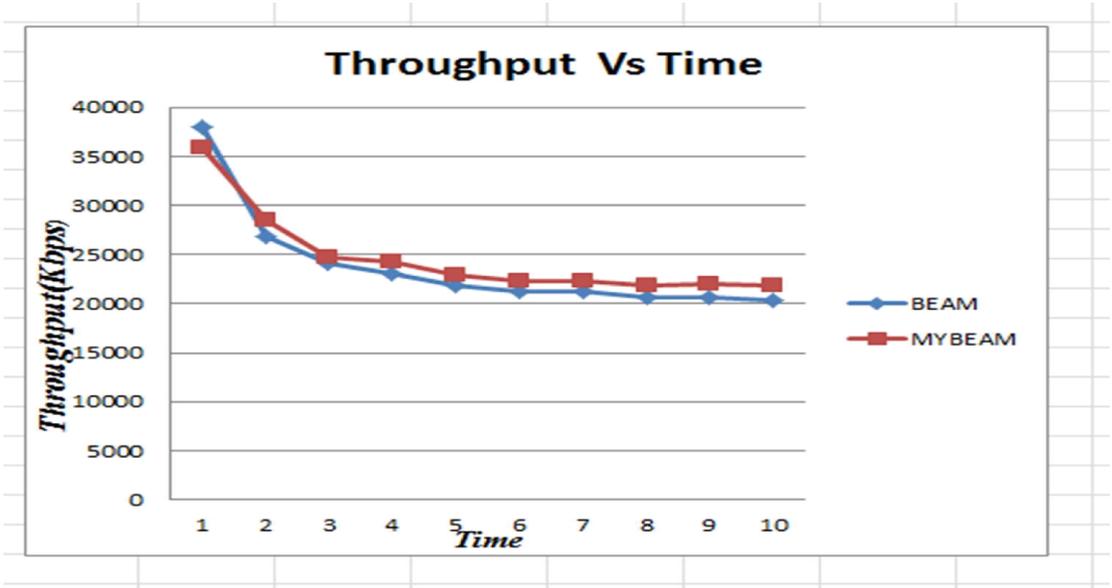

Figure 9 Throughput via Simulation Time

Figure 9 illustrates, the enhanced performance of the cluster-based MYBEAM Protocol in terms of throughput. The graph indicates that the MYBEAM Protocol exhibits higher throughput compared to the existing BEAM Protocol. The X-axis represents simulation time, while the Y-axis displays throughput measured in kilobits per second (kbps). Initially, a slight variation is observed between 1 and 2 seconds, it might be due to the cluster formation process and the joining time of vehicles outside the communication range of Roadside Units (RSUs). This period may also reflect the prediction of emergency situations by RSUs and Cluster Heads (CHs), thereby alerting all cluster heads and members. Finally, the throughput increases significantly when compared to the BEAM Protocol, which lacks clusters and vehicle-to-vehicle (V2V) communication; MYBEAM shows a gradual increase it throughput after 3 seconds.

**Packet Delivery Ratio:** Data Delivery Ratio is the metrics used to measure the percentage of data message which are successfully received by vehicles over a network. This illustrates the level of delivered data to the destination. It is calculated by the number of received packets by the number of sent packet multiply by 100

$$PDR = \frac{(Number\ of\ Recived\ packets)*(100)}{Number\ of\ sent\ packets} \quad\dots\dots\dots\dots\dots\dots\dots\dots\dots\dots\dots\dots\dots\dots\dots\dots (12)$$

The greater value of packet delivery ratio means the better performance of the protocol.

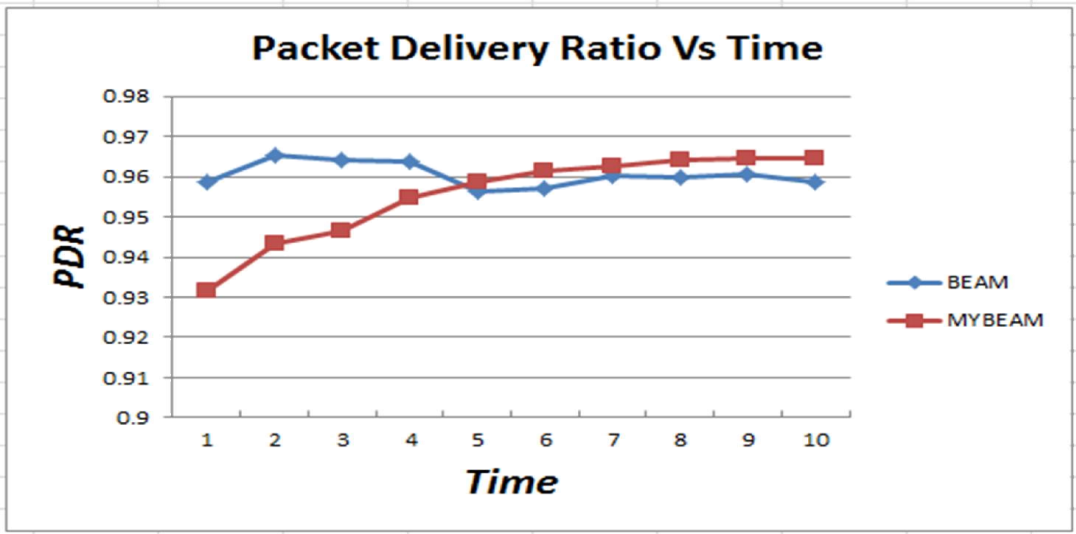

Figure 10 Packet Delivery Ratio via Simulation Time

Figure 10 indicates that the MYBEAM Protocol exhibits a higher Packet Delivery Ratio (PDR) compared to the BEAM Protocol. However, between 1 and 5 seconds, the BEAM Protocol demonstrates superior performance, as it did not involve additional processing such as cluster formation or the transmission of extra routing packets. After 5 seconds, MYBEAM outperforms BEAM due to its stable clustering in a dynamic network topology, facilitated by the selection of a Second Cluster Head, which helps prevent re-clustering.

**Average End-to-End Delay**

**End to end Delay** is defined as the difference between the time a sender generates a packet and the time the receiver receives it. This metric represents the average time required for a packet to reach its destination and encompasses various sources of delay, including re-clustering, route discovery, queuing, retransmission delays, and packet drops [36].

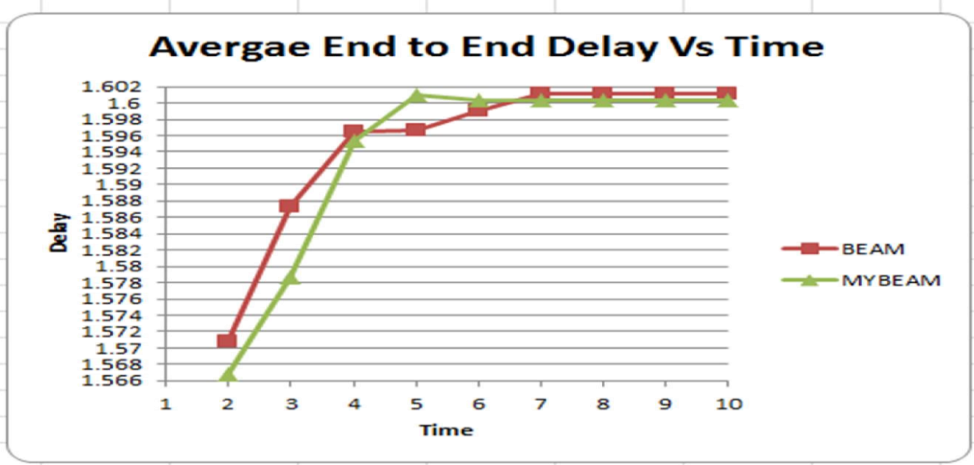

Figure 11 Average End to End Delay via Time

Figure 11 illustrates the improved performance of the MYBEAM Protocol in terms of end-to-end delay. Between 1 and 4 seconds, MYBEAM exhibits lower delay compared to the BEAM Protocol. However, from 4 to 6 seconds, the delay for MYBEAM exceeds that of BEAM due to the re-clustering process; packets may be sent but do not reach their destination during this period. After 7 seconds, the delays for both MYBEAM and BEAM stabilize and become constant.

**Conclusion and Future Work**

This paper, has developed an architectural algorithm for stable cluster communication and extended link longevity in VANETs, addressing multicar chain collisions associated with the BEAM protocol. While BEAM enhances performance by predicting emergencies, it only informs multicast group members, leaving ordinary vehicles unaware, which can lead to collisions. To resolve this, we have proposed a multi-agent-based stable clustering technique using roadside unit agents and vehicle agents. This algorithm adjusts cluster sizes based on vehicle mobility and selects cluster heads based on weight factors to minimize re-clustering and ensure ordinary vehicles are notified of emergencies. When an emergency is detected, the RSU sends an Emergency Warning Message (EWM) to cluster heads, who relay it to their members. If a CH from ordinary vehicles identifies an emergency, it informs the multicast group heads, and then the CH informs RSU and the RSU informs to all CH and the CH Informs to all members and other

CH, to prevent collisions. Simulations in NS-2 with SUMO-0.26.0 show that the proposed method effectively improves throughput, packet delivery ratio, and end-to-end delay across varying simulation times.

This research presents several suggestions for further studies. The current implementation utilizes additional routing packets sent by cluster heads to discover vehicles outside the roadside unit's communication range, which increases network overhead as simulation time extends. Future efforts should aim to reduce this overhead to stabilize communication. Additionally, some vehicles do not respond to join control packets even within the roadside unit's communication range, often due to link failures. As this research is based on simulations, further validation in real-world environments may uncover additional challenges not addressed here. Moreover, while the current study focuses on text data for disseminating emergency information, future work should explore support for real-time data and quality of service for traffic types such as video, audio, and other entertainment formats in vehicular ad hoc networks.

**Funding Declaration: No funding**

*Biography*

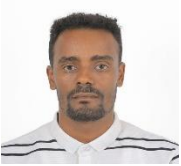
*Bsc In Information Technology in Jimma University in 2016 , MSc In computer Network from Jimma University in 2020, working as lecturer and Network Administrator in Jimma University* **[1]**

*Bsc In Computer Science in Jimma University in 2010 , MSc In computer Network from Jimma University in 2018, PhD In Computer networking from Addis Ababa University* **[2]**

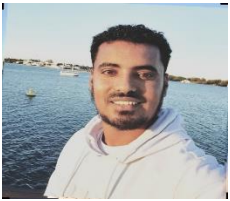
*Bsc In Information Technology in Jimma University in 2015 , MSc In computer Network from Jimma University in 2018, PhD Student in Griffith University* **[3]**